\documentclass[%
superscriptaddress,
twocolumn,
prl,
raggedbottom,
]{revtex4-2}

\usepackage{graphicx}% Include figure files
\usepackage{dcolumn}% Align table columns on decimal point
\usepackage{bm}% bold math
\usepackage{amsmath,amsthm,amssymb}
\usepackage[separate-uncertainty = true]{siunitx} %[separate-uncertainty = true,multi-part-units=single]
\usepackage{color}
\DeclareSIUnit\permille{\text{\textperthousand}}
\usepackage[english]{babel}
\usepackage{amsfonts}
\usepackage{appendix}
\usepackage{textcomp} % to use \textperthousand 
\definecolor{blue(ncs)}{rgb}{0.0, 0.53, 0.74}
\begin{document}

\preprint{APS/123-QED}

\title{All-optical GeV electron bunch generation in a laser-plasma accelerator via truncated-channel injection \\
}% Force line breaks with \\

\author{A. Picksley}
\email{apicksley@lbl.gov}
\altaffiliation[Present address: ]{Lawrence Berkeley National Laboratory, Berkeley, California 94720, USA}
\affiliation{John Adams Institute for Accelerator Science and Department of Physics, University of Oxford, Denys Wilkinson Building, Keble Road, Oxford OX1 3RH, United Kingdom}%
\author{J. Chappell}%
\affiliation{John Adams Institute for Accelerator Science and Department of Physics, University of Oxford, Denys Wilkinson Building, Keble Road, Oxford OX1 3RH, United Kingdom}%
\author{E. Archer}%
\affiliation{John Adams Institute for Accelerator Science and Department of Physics, University of Oxford, Denys Wilkinson Building, Keble Road, Oxford OX1 3RH, United Kingdom}%
\author{N. Bourgeois}%
\affiliation{Central Laser Facility, STFC Rutherford Appleton Laboratory, Didcot OX11 0QX, United Kingdom}%
\author{J. Cowley}%
\affiliation{John Adams Institute for Accelerator Science and Department of Physics, University of Oxford, Denys Wilkinson Building, Keble Road, Oxford OX1 3RH, United Kingdom}%
\author{D. R. Emerson}
\affiliation{Scientific Computing Department, STFC Daresbury Laboratory, Warrington WA4 4AD, United Kingdom}
\author{L. Feder}%
\affiliation{John Adams Institute for Accelerator Science and Department of Physics, University of Oxford, Denys Wilkinson Building, Keble Road, Oxford OX1 3RH, United Kingdom}%
\author{X. J. Gu}
\affiliation{Scientific Computing Department, STFC Daresbury Laboratory, Warrington WA4 4AD, United Kingdom}
\author{O. Jakobsson}%
\affiliation{John Adams Institute for Accelerator Science and Department of Physics, University of Oxford, Denys Wilkinson Building, Keble Road, Oxford OX1 3RH, United Kingdom}%
\author{A. J. Ross}%
\affiliation{John Adams Institute for Accelerator Science and Department of Physics, University of Oxford, Denys Wilkinson Building, Keble Road, Oxford OX1 3RH, United Kingdom}%
\author{W. Wang}%
\affiliation{John Adams Institute for Accelerator Science and Department of Physics, University of Oxford, Denys Wilkinson Building, Keble Road, Oxford OX1 3RH, United Kingdom}%
\author{R. Walczak}%
\affiliation{John Adams Institute for Accelerator Science and Department of Physics, University of Oxford, Denys Wilkinson Building, Keble Road, Oxford OX1 3RH, United Kingdom}%
\affiliation{Somerville College, Woodstock Road, Oxford OX2 6HD, United Kingdom}
\author{S. M. Hooker} 
\email{simon.hooker@physics.ox.ac.uk}
\affiliation{John Adams Institute for Accelerator Science and Department of Physics, University of Oxford, Denys Wilkinson Building, Keble Road, Oxford OX1 3RH, United Kingdom}%

\date{\today}% It is always \today, today,
       % but any date may be explicitly specified

\begin{abstract}
We describe a simple scheme, truncated-channel injection, to inject electrons directly into the wakefield driven by a drive pulse guided by an all-optical plasma channel. We use this approach to generate dark-current-free $\SI{1.2}{GeV}$, 4.5\% relative energy spread electron bunches with \SI{120}{TW} laser pulses guided in a \SI{110}{mm}-long hydrodynamic optical-field-ionized (HOFI) plasma channel. Our experiments and particle-in-cell simulations show that high-quality electron bunches were only obtained when the drive pulse was closely aligned with the channel axis, and was focused close to the density down-ramp formed at the channel entrance. Start-to-end simulations of the channel formation, and electron injection and acceleration show that increasing the channel length to \SI{410}{mm} would yield \SI{3.65}{GeV} bunches, with a slice energy spread $\sim \SI{5E-4}{}$.

This article was published in Physical Review Letters 131, 245001 on 12 December 2023. DOI: \url{https://doi.org/10.1103/PhysRevLett.131.245001}

\copyright 2023 American Physical Society

\end{abstract}

\maketitle

It has become well established that laser-plasma accelerators (LPAs) \cite{Dawson1979} can accelerate few-femtosecond \cite{vanTilborg.2006, Ohkubo.2007, Debus.2010, Lundh2011,Heigoldt.2015} electron bunches to GeV-scale energies in accelerator stages only a few centimetres long \cite{Leemans2006, Karsch2007, Wang2013, Leemans2014, Shin.2018, Gonsalves2019, Ke2021}. These highly desirable features make LPAs attractive for driving compact, femtosecond-duration light sources \cite{Corde2013,albert2016app}, including free-electron lasers (FELs). FEL gain has recently been reported in experiments utilizing laser- \cite{Wang2021, Labat.2022nd8} and beam-driven \cite{Pompili.2022, Galletti.2022} plasma accelerators. Other work has demonstrated generation of incoherent soft x-radiation in an undulator \cite{Fuchs2009}, and at photon energies in the keV range from betatron oscillations \cite{Kneip.2010}, and at MeV energies from Thomson scattering \cite{Phuoc.2012, Powers.2013, Khrennikov.2015}.

These demanding applications require generation of multi-GeV electron bunches with high peak current, low transverse emittance, and small shot-to-shot jitter of the bunch properties. Therefore, it is preferable to operate in the linear, or quasi-linear, regime \cite{Esarey.2009} in order to prevent uncontrolled self-injection at one or more points along the length of the accelerator. However, this regime brings two significant challenges. First, relativistic self-focusing does not occur and hence the drive laser pulse must be guided over several Rayleigh lengths to reach multi-GeV energies. Second, trapping of electrons is more difficult since the electric fields of the plasma wave are lower than in the highly nonlinear regime \cite{Esarey.2009}.

In order to meet the first of these challenges, we have developed low-density hydrodynamic optical-field-ionized (HOFI) plasma channels \cite{Gonsalves.2017,robert2018a, Shalloo2018, Shalloo2019}, building on pioneering work by Milchberg et al. \cite{Durfee1993, Durfee1995, Clark1997, Kumarappan2005}. These channels have low losses, and could in principle be several metres long \cite{Picksley2020a}. Experiments have demonstrated guiding of relativistically-intense laser pulses through \SI{200}{mm}-long channels with axial densities as low as $\SI{1E17}{cm^{-3}}$ \cite{Picksley2020, Picksley2020a, Miao2020, Feder2020}, and operation at kHz pulse repetition-rates \cite{Alejo2022}. 

In this Letter we demonstrate a new approach that addresses both challenges in a single stage. We show that the density down-ramp formed at the start of a HOFI channel promotes electron injection directly into the quasi-linear wakefield driven by a channel-guided drive pulse. We demonstrate experimentally that this truncated-channel injection (TCI) scheme produces electron bunches with an energy up to \SI{1.2}{GeV}, and a root-mean-square (rms) energy spread of $\sigma_\mathrm{E} / \mu_\mathrm{E} = \SI{4.5}{\%}$ with \SI{120}{TW} laser pulses. We show that high-quality TCI bunches are only generated when the drive laser is: (i) focused close to the down-ramp at the channel entrance; and (ii) well-aligned with, and hence guided by, the HOFI channel. In contrast, we find that bunches produced by ionization injection have much larger energy spread and are preferentially generated when the drive pulse is mis-aligned with the channel. The experimental results are found to be in excellent agreement with start-to-end simulations of the complete TCI scheme, which includes hydrodynamic simulations of the formation of the HOFI channel and high-resolution particle-in-cell (PIC) simulations of electron injection and acceleration. These simulations show that dephasing was not reached in our experiments, and that extending the channel length to the dephasing length $L_\mathrm{d} \approx \SI{410}{mm}$ would yield bunches of energy \SI{3.65}{GeV}, a slice energy spread below the per-mille level, a peak current of \SI{0.8}{kA}, and a normalized transverse emittance of $\epsilon_{n,\perp} < \SI{5}{mm.mrad}$. These properties appear well-matched to the requirements of soft X-ray FELs driven by \SI{100}{TW} class lasers.

Before describing this work in detail we note that Scott et al.\ \cite{Scott2020} investigated electron injection in a pre-pulse generated plasma structure similar to that investigated here, but not coupled into a waveguide. Oubrerie et al. generated \SI{1.1}{GeV} bunches with relative energy spread $\SI{4}{\%}$, following blade-induced shock injection in a higher-density, 15-mm-long, all-optical channel \cite{Oubrerie2022}. Miao et al. used \SI{300}{TW} laser pulses to generate $\SI{5}{GeV}$ bunches with large energy spreads following ionization injection in nitrogen-doped, 20-cm-long HOFI channels \cite{Miao2022}. Recent work investigating density transitions generated by hydrodynamically-expanding shocks has demonstrated their feasibility for beam-driven \cite{Foerster2022} and laser-driven \cite{v2023laser} acceleration.

\begin{figure}[t!]
  \centering
  \includegraphics[width=\linewidth]{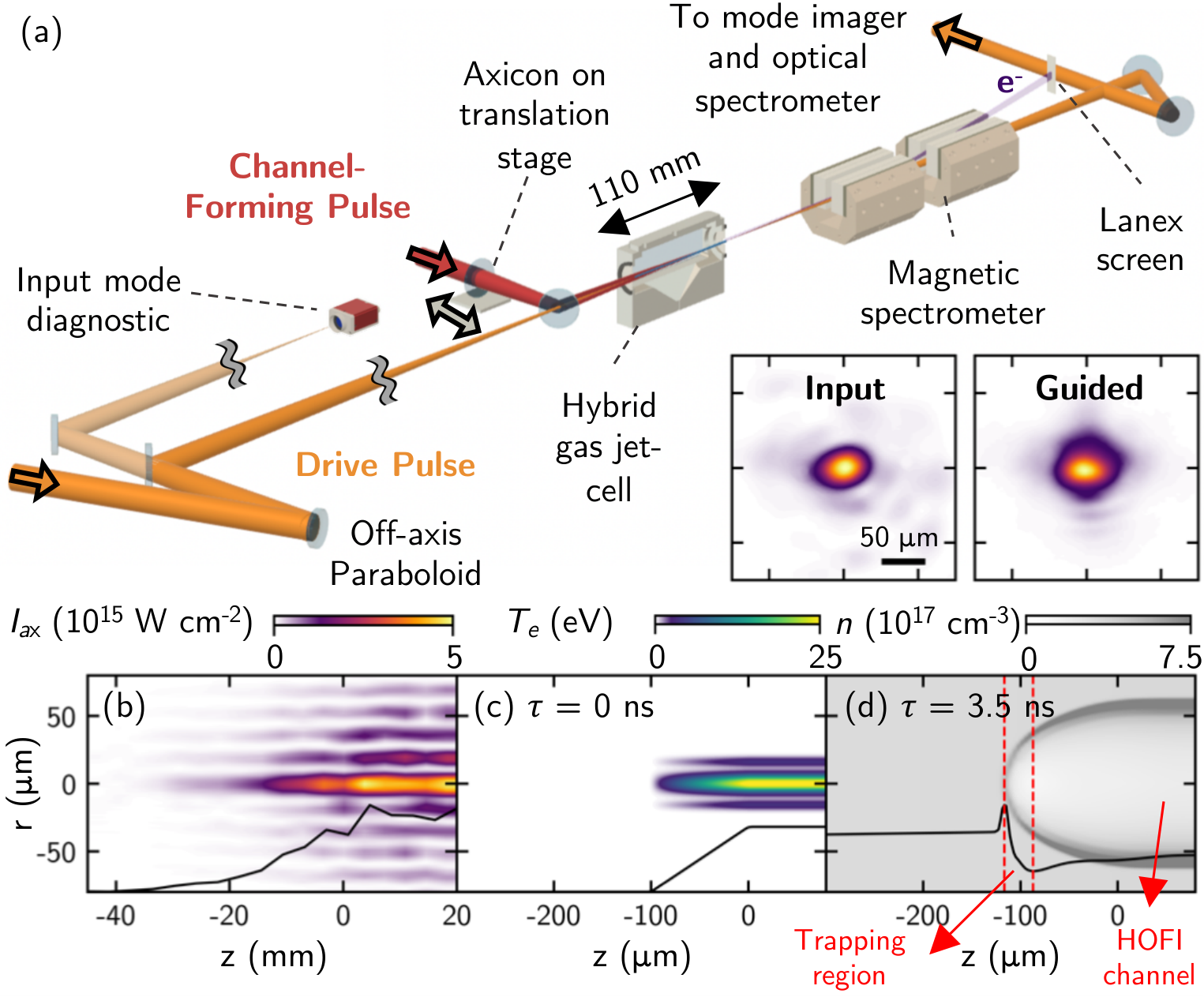}
  \caption{Schematic of truncated-channel injection scheme. (a) Setup: The channel-forming (red) and drive (orange) beams were coupled into the gas target. The input mode, output mode, optical and electron spectra were measured on every shot. Inset: Measured transverse fluence profiles of the drive laser at focus and at the exit of the HOFI channel. (b) Measured axicon longitudinal intensity profile, (c) calculated initial electron temperature profile, and (d) calculated density profile of the truncated HOFI plasma channel $\SI{3.5}{ns}$ after arrival of the channel-forming pulse. In each panel, the black curve shows the relative magnitude of each variable along the optical axis. 
  }
  \label{fig:setup}
\end{figure}

Figure \ref{fig:setup} outlines the TCI scheme, and the experimental arrangement employed using the Astra-Gemini TA3 Ti:sapphire laser at the Rutherford Appleton Laboratory which delivers $\SI{47(3)}{fs}$ pulses with $\lambda_0 = \SI{800}{nm}$. The channel-forming beam, of pulse energy \SI{86(17)}{mJ}, was focused by an axicon lens to form a plasma column within the 110-mm-long gas target. The drive beam, containing \SI{5.8(0.2)}{J}, was focused into the gas target to a spot-size $w_0 \approx \SI{40}{\micro m}$, with peak intensity $I_\mathrm{pk} \approx \SI{2.4e18}{W.cm^{-2}}$ ($a_0 \approx 1.0$). The delay between the arrival of the two pulses was set to $\Delta \tau = \SI{3.5}{ns}$. The target comprised a \SI{110}{mm} long hybrid gas jet-cell \cite{Aniculaesei2018} filled with a \SI{2}{\%} mix of nitrogen in hydrogen. 

A small fraction ($< \SI{1}{\%}$) of the drive laser was transmitted through the final turning mirror and focused in order to measure on each shot the input transverse offset, $\delta r$, of the drive pulse focus relative to the channel. After leaving the cell, the drive pulse was directed to a far-field camera, and two fiber-based optical spectrometers. Electron beam characteristics were measured by an electron spectrometer.

Truncating the plasma channel to trigger injection was achieved by delaying the longitudinal position of the start of the HOFI plasma channel $z_\mathrm{ch}$ as follows. In principle $z_\mathrm{ch}$ is calculated from the diameter (\SI{26}{mm}) of the central hole in the axicon and the approach angle of the axicon ($1.6^{\circ}$). In practice, however, wavefront non-uniformities and diffraction determine the distance over which the intensity of the line focus increases from zero to the threshold intensity for ionisation. Figure \ref{fig:setup}(b) shows the measured intensity distribution, $I_{\mathrm{ax}}(r,z)$, near the start of the axicon line focus, and Fig.~\ref{fig:setup}(c) shows the calculated electron temperature profile $T_\mathrm{e}(r,z)$ of the resulting initial plasma column~\cite{supp_mat}. The strong intensity dependence of optical field ionization causes the onset of the plasma column to be significantly sharper than that of $I_{\mathrm{ax}}(r,z)$. Hydrodynamic simulations [see Fig. \ref{fig:setup}(d)] show that the front edge of the plasma column expands longitudinally to form a hemispherical acoustic shock wave, resembling a Sedov-Taylor \textit{spherical} expansion of electron and neutral density ($n = n_\mathrm{e} + n_\mathrm{H}$) \cite{Hutchens2000, supp_mat}. The body of the plasma column expands radially, driving a \textit{cylindrical} shock into the surrounding gas \cite{Shalloo2018, Picksley2020a}. The leading edge of the drive laser will ionize any neutral atoms to give a new electron density equal to $n$ \cite{Picksley2020a, Feder2020}. The scale length of the transition, $\hat{L}_\mathrm{tr} = n_\mathrm{e0} / (d n_\mathrm{e} / d z )$, where $n_\mathrm{e0}$ is the axial density in the HOFI channel, is varied by adjusting the position of $z_\mathrm{ch}$ within the plume of gas exiting the cell entrance.Since the expansion rates of the hemi-spherical and cylindrical blast-waves are different, independent control over the parameters of the injection and acceleration section can be achieved (see Appendix). The calculated $\hat{L}_\mathrm{tr} \approx \SI{15}{\micro m}$ is short compared to the plasma wavelength $\lambda_p = 2 \pi c (n_{e0} e^2 / m_e \epsilon_0)^{-1/2} \approx \SI{100}{\micro m}$, hence localized injection is triggered by an abrupt change in wakefield phase as the plasma electrons cross the density transition \cite{Suk2001, Schmid2010, Buck2013}. 

We examined the TCI scheme by setting the longitudinal position of the drive pulse focus, $z_\mathrm{f}$, to coincide with the position of the front pinhole, and varying $z_\mathrm{ch}$ so as to adjust their separation, $\Delta z = z_\mathrm{ch} - z_\mathrm{f}$, between \SI{-11.2}{mm} and \SI{+0.8}{mm}. We note that in a standard guiding experiment $\Delta z \ll \SI{0}{mm}$, so that the entrance to the channel is far upstream of the drive pulse focus. The axial plasma density was set to $\sim \SI{1.3(0.1)e17}{cm^{-3}}$, suppressing ionization injection for guided pulses~\cite{supp_mat} when $\Delta z = \SI{-11.2}{mm}$.

\begin{figure}[!t]
  \centering
  \includegraphics[width=0.48\textwidth]{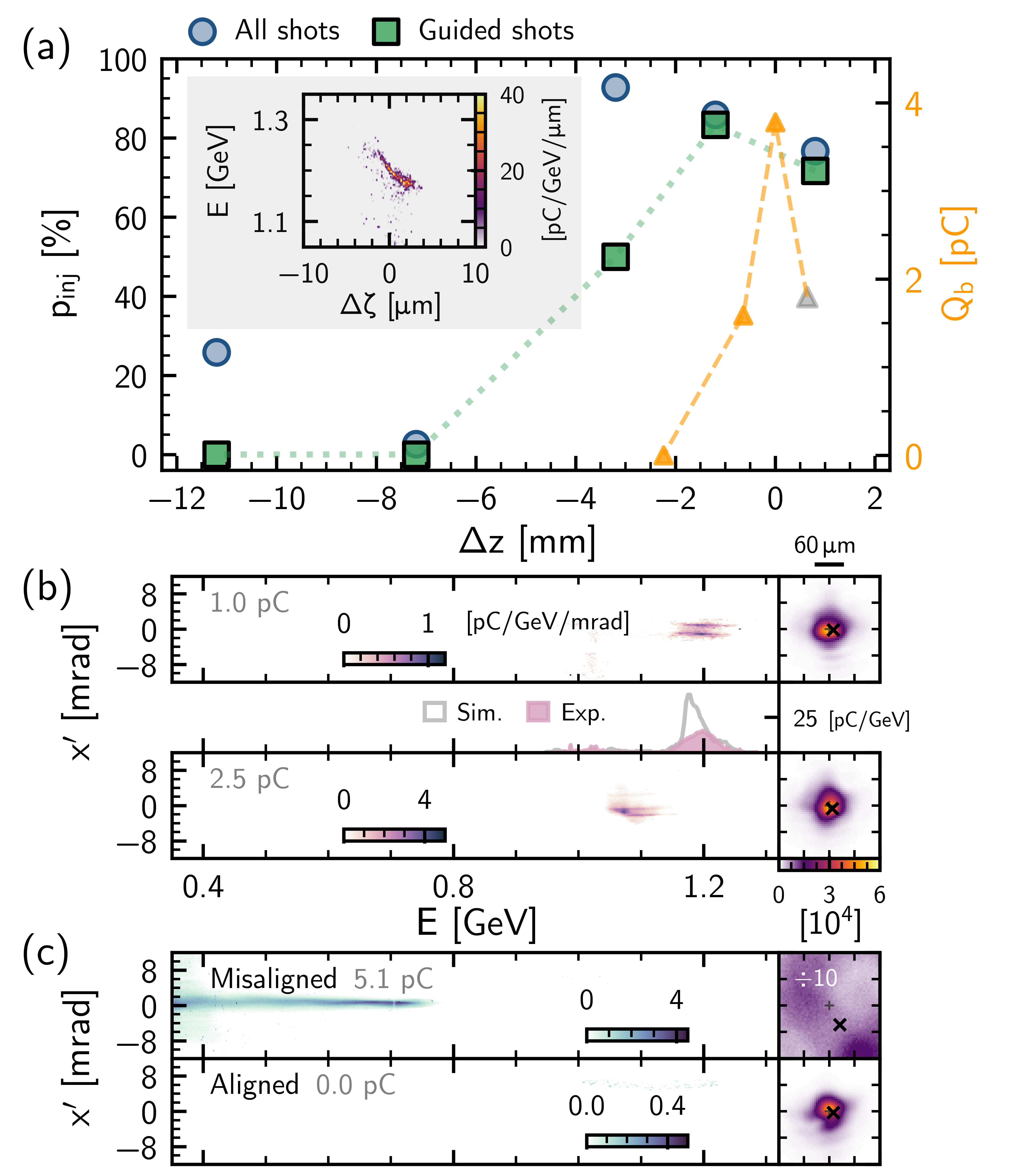}
  \caption{(a) Left axis: Injection probability, $p_\mathrm{inj}$, as a function of $\Delta z = z_\mathrm{ch} - z_\mathrm{f}$ for all (blue circles) and guided (green squares) shots. Right axis: Variation of calculated electron bunch charge $Q_b$ with $\Delta z$, from PIC simulations. Inset (grey): Simulated longitudinal phase-space distribution of the electron bunch at the channel exit for $\Delta z = \SI{0.8}{mm}$. (b) Examples of the measured, angularly-resolved, electron spectra (left), and guided drive beam profile (right) for $\Delta z = \SI{0.8}{mm}$ --- black crosses show the drive input position. Also shown is a comparison between the measured and simulated spectral density. (c) Plots as in (b) for $\Delta z = \SI{-11.2}{mm}$, for cases where the drive pulse was mis-aligned (upper) or aligned (lower) to the channel.}
  \label{fig:TCIresults}
\end{figure}

Figure \ref{fig:TCIresults}(a) shows the recorded injection probability, $p_{\mathrm{inj}}$, as a function of $\Delta z$, where injection is considered to have occurred if the total charge recorded by the electron spectrometer exceeded \SI{0.05}{pC}. For guided shots, electron bunches were observed for $\SI{-3.2}{mm} \leq \Delta z \leq \SI{0.8}{mm}$, i.e.\ when the the drive focus was well within one Rayleigh length of the leading edge of the plasma channel. For 124 guided shots with $\Delta z \geq \SI{-3.2}{mm}$, $p_{\mathrm{inj}} = 74\%$, whereas no electron injection was observed in 73 guided shots when $\Delta z \leq \SI{-7.2}{mm}$. With the channel entrance at its most downstream position, $\Delta z = \SI{0.8}{mm}$, high-quality guiding and electron acceleration were observed simultaneously when the input beam was well-aligned with the HOFI channel, as shown in Fig.~\ref{fig:TCIresults}(b). Electron bunches of pC-scale charge with mean energies in excess of \SI{1}{GeV} and few-percent energy spreads were consistently observed for input offsets $\delta r \lesssim \SI{10}{\micro m}$. These shots exhibited an angular splitting of the bunch; PIC simulations indicated that this arose from transverse oscillations of the bunch that are seeded by laser mode beating \cite{supp_mat}. When the channel entrance was upstream of the drive focus ($\Delta z \leq \SI{-7.2}{mm}$), electron injection was not observed for guided shots, and instead injection was only observed when the drive pulse was offset transversely from the channel axis [see Fig.~\ref{fig:TCIresults}(c)]. 

\begin{figure}[t!]
  \centering
  \includegraphics[width=0.48\textwidth]{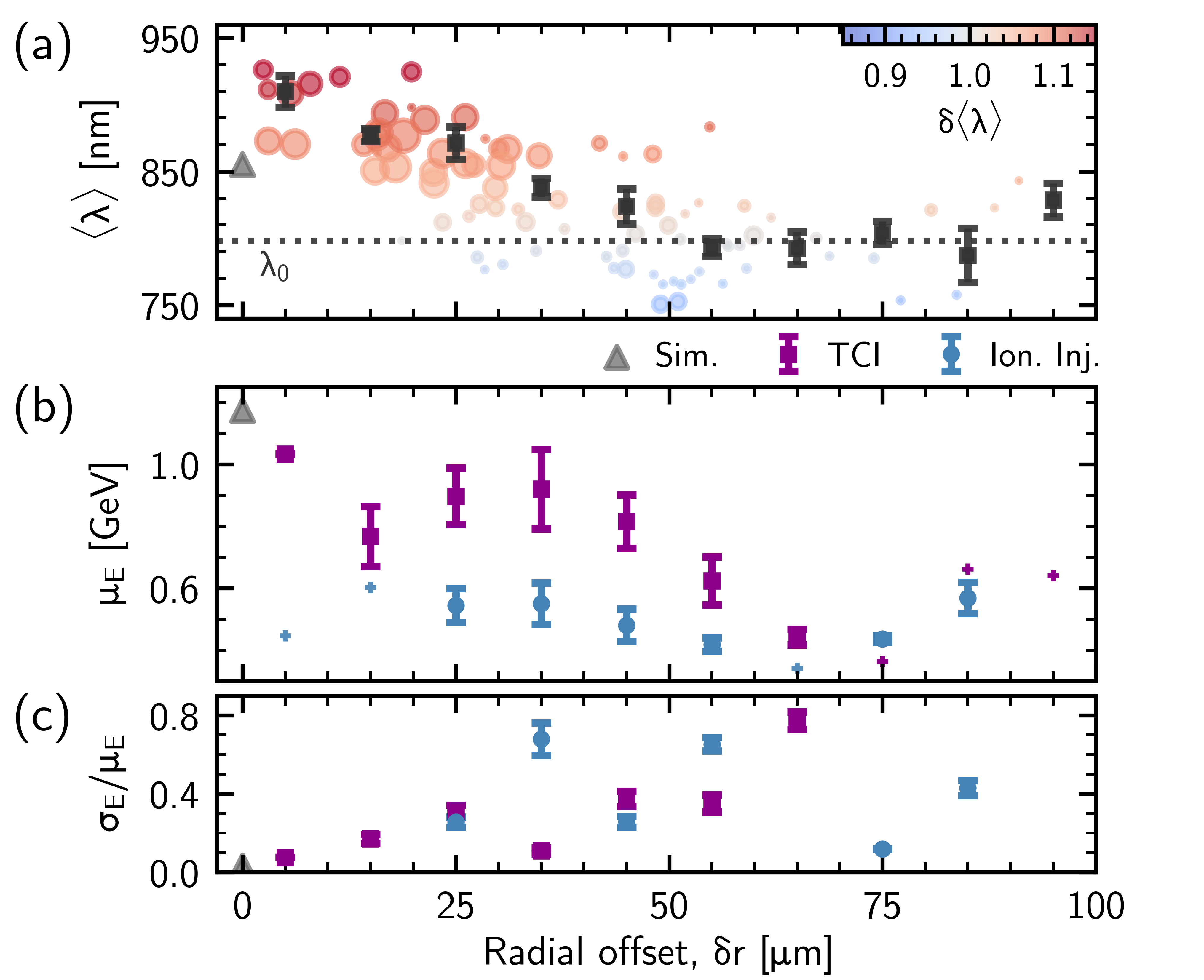}
  \caption{Variation with $\delta r$ of the spectra of the output drive and electron beams. (a) $\langle \lambda \rangle$ for $\Delta z = \SI{0.8}{mm}$. The size of data points corresponds to the relative fluence of the measured guided mode. The result from start-to-end simulations is shown in gray. (b) Mean electron energy, $\mu_E$, for guided shots for TCI (purple squares) and ionization injection (blue circles); bins containing only a single event use `$+$' markers. For the ionization injection dataset, $\Delta z = \SI{-11.2}{mm}$ and $n_{e0} = \SI{2.2(0.1)e17}{cm^{-3}}$. (c) Average ratio $\sigma_E / \mu_E$ of accelerated electron bunches. For all plots the data points with error bars are averages over \SI{10}{\micro m}-wide bins~\cite{supp_mat}, with the error bars showing one standard error.}
  \label{fig:TCIspectra}
\end{figure}

The spectral shift of the drive pulses provides insight into wakefield excitation along the plasma channel. Figure \ref{fig:TCIspectra}(a) shows, as a function of $\delta r$, the intensity-weighted average wavelength, $\langle \lambda \rangle$, of the transmitted drive pulse, together with a measure of the wavelength shift, $\delta \langle \lambda \rangle = \langle \lambda \rangle / \lambda_0$. Well-aligned shots ($\delta r \lesssim \SI{30}{\micro m}$) are strongly correlated with high transmission and significant red-shifting, corresponding to good guiding and strong wakefield excitation. In contrast, shots with large $\delta r$ are strongly associated with low transmission and significant blue-shifting, consistent with significant ionization by the drive pulse. 

Figures~\ref{fig:TCIspectra}(b) and (c) show corresponding effects in the measured electron energies, with larger mean energies $\mu_E$ measured for smaller $\delta r$. As shown in Fig.\ \ref{fig:TCIspectra}(c), well-aligned shots ($\delta r \leq \SI{10}{\micro m}$) have dramatically lower relative rms energy spread --- $\sigma_E / \mu_E \approx 7\%$ on average, with a best of $4.5\%$ --- than less well-aligned shots. Furthermore, well-aligned shots exhibited percent-level shot-to-shot energy stability, with a mean energy of $\mu_{\mathrm{E}} = \SI{1.033(0.010)}{GeV}$. For larger input offsets ($10 < \delta r \leq \SI{50}{\micro m}$), electron bunches of higher charge ($\sim \SI{10}{pC}$) were obtained with spectra typically comprising multiple peaks superposed on a continuous spectrum extending to $\lesssim\SI{1.0}{GeV}$, with a mean energy $\mu_\mathrm{E} \sim \SI{0.8}{GeV}$. For $\delta r > \SI{50}{\micro m}$, electron spectra were effectively continuous, with $\mu_{\mathrm{E}} \sim \SI{0.5}{GeV}$. 

Figures~\ref{fig:TCIspectra}(b) and (c) also show results for the case of ionization injection from the nitrogen dopant, which was studied by setting $n_\mathrm{e0} = \SI{2.2(0.1)e17}{cm^{-3}}$ and translating the axicon to $\Delta z = \SI{-11.2}{mm}$. It is noticeable that for ionization injection,  $\mu_E$ is lower and much less sensitive to $\delta r$. Electrons were preferentially injected when the drive pulse was slightly mis-aligned with respect to the axis of the channel. Injection occurred only twice in 26 shots with $\delta r \leq \SI{20}{\micro m}$ ($\sim 7\%)$ but occurred in $\sim 30\%$ of guided events for $20 < \delta r \leq \SI{50}{\micro m}$.

\begin{figure}[t!]
  \centering
  \includegraphics[width=\linewidth]{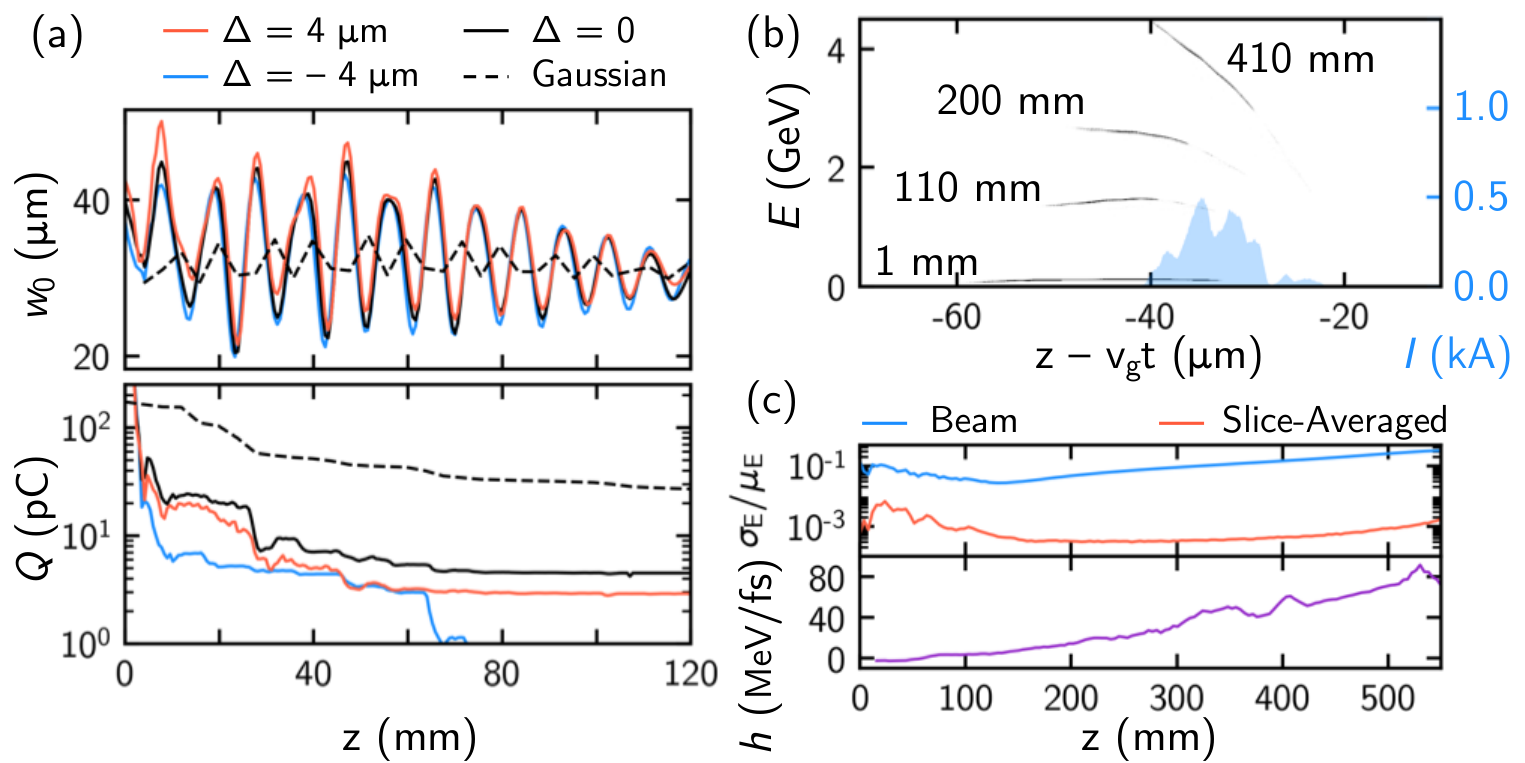}
  \caption{(a) Laser spot size (top) and TCI bunch charge (bottom) during propagation for input modes of different sizes. (b) Electron bunch longitudinal phase-space at four different propagation distances. The longitudinal current profile at $z = \SI{410}{mm}$ is shown in blue. (c) Relative energy spread of the TCI bunch ($\sigma_E/\mu_E$, blue), the mean relative slice energy spread (red), and bunch chirp ($h$, purple).}
  \label{fig:simulations}
\end{figure}

The data presented in Figs. \ref{fig:TCIresults} and \ref{fig:TCIspectra} may be interpreted as follows. At the lower density investigated in this work, high quality electron injection only occurred when the drive pulses were: (i) well-aligned with the channel ($\delta r < \SI{10}{\micro m}$); and (ii) focused close to the channel entrance ($\Delta z = \SI{0.8}{mm}$). For these conditions, strong wakefield excitation occured, and $\sim \SI{1}{GeV}$ electron bunches with few-percent energy spread were generated with an injection probability of $\sim 80\%$. When the channel entrance was located far upstream of the drive focus, electron bunches were only generated if the drive pulse was mis-aligned; in this case, the electron spectra were broad-band, and the transmitted drive pulses had low energy transmission with much reduced red-shifting. These observations are consistent with poorly-aligned drive pulses interacting with the higher-density channel walls and surrounding neutral gas, triggering uncontrolled electron injection, potentially at several points along the channel. For the case of ionization injection, broad-band electron bunches were generated for a wide range of $\delta r$, suggesting that high-quality guiding was not necessary, and that injection occurred at several points along the plasma channel.

To gain further insight we performed start-to-end modelling of the channel formation and electron acceleration process \cite{supp_mat}. An Airy drive laser focus with $\delta r = 0$ and other parameters similar to the experiment was assumed. The right-hand axis of Fig.~\ref{fig:TCIresults}(a) shows the calculated electron bunch charge $Q_b$ at the channel exit for several values of $\Delta z$. The variation of $Q_b$ with $\Delta z$ is qualitatively similar to $p_\mathrm{inj}$, although the range of $\Delta z$ over which injection is observed in simulation is smaller. The inset to this Figure shows the calculated longitudinal phase space of the bunch at the exit of the channel for $\Delta z = \SI{0.8}{mm}$. The properties of this bunch are: $\mu_\mathrm{E} = \SI{1.175}{GeV}$, $\sigma_E = \SI{38}{MeV}$ ($\SI{3.2}{\%}$), and $Q_b = \SI{1.8}{pC}$. The calculated FWHM bunch duration was \SI{9.4}{fs}, and the mean slice energy spread was $\SI{1.0(0.4)}{\%}$. The spectra of the electron bunch and the transmitted drive pulse are found to be in good agreement with the measurements [see Figs.~\ref{fig:TCIresults}(b) and~\ref{fig:TCIspectra}(a)].

Efficient coupling of the drive pulse and electron bunch into the HOFI channel depended on $w_0$ and $z_f$. For the conditions of Fig.~\ref{fig:TCIresults}, approximately $\SI{20}{\%}$ of the trapped charge was coupled into the channel, the rest being immediately dephased. Rapid spot-size oscillations, caused by improper matching of the drive pulse into the HOFI channel, led to transverse ejection of particles from the wakefield. Figure~\ref{fig:simulations}(a) shows the evolution with $z$ of the mode size $w(z)$ and $Q_b$ for the same channel parameters as Fig.~\ref{fig:TCIresults} and several Airy input modes mismatched to the channel by $\Delta = w_\mathrm{m} - w_0$. For a Gaussian mode with $w_0 \approx w_\mathrm{m}$, $> \SI{40}{\%}$ of injected bunch charge is transported into the channel, increasing the peak current by an order of magnitude compared to the Airy mode.

Figure~\ref{fig:simulations} also shows the evolution of the bunch characteristics for channels with a length up to the dephasing length $L_\mathrm{d} \approx \SI{410}{mm}$, i.e. longer than investigated here, but experimentally feasible~\cite{Picksley2020a,Feder2020}. At $z = L_\mathrm{d}$, the properties of the bunch are found to be: $\mu_\mathrm{E} = \SI{3.65}{GeV}$; $\sigma_\mathrm{E} = \SI{487}{MeV}$ ($\sigma_\mathrm{E} / \mu_E = \SI{13.3}{\%}$); an rms duration of $\sim \SI{26}{fs}$, corresponding to a peak current $I_\mathrm{peak} \approx \SI{0.7}{kA}$; and a normalized projected transverse emittance of $\epsilon_{n,\perp} \approx \SI{5.8}{mm.mrad}$ (with a slice-average of $\epsilon_{n,\perp}^\mathrm{slice} \approx \SI{3.6}{mm.mrad}$). 

Evolution of the bunch chirp, $h(z)$, is shown in Fig.~\ref{fig:simulations}(c). Upon injection, TCI bunches exhibit negative chirp ($h \approx \SI{-1.2}{MeV.fs^{-1}}$), during propagation the accelerating field causes $h$ to increase to $\approx \SI{49}{MeV.fs^{-1}}$. Thanks to the linear longitudinal phase-space distribution, a plasma-based dechirper~\cite{d2019tunable} could reduce $\sigma_\mathrm{E}$ to the level of the slice-averaged energy spread, $\sigma_\mathrm{E}^{\mathrm{slice}}/\mu_\mathrm{E} \sim \SI{5.e-4}{}$, potentially producing an FEL-quality bunch. In the 1D limit, three bunch conditions must be satisfied \cite{huang2007review}, of which we find the most stringent is: $\epsilon_\perp = \epsilon_{n,\perp}^\mathrm{slice} / \gamma < \lambda_R / 4 \pi$, where $\gamma$ is the Lorentz factor and $\lambda_R$ is the FEL wavelength. This results in $\lambda_R \gtrsim 4 \pi \epsilon_\perp \approx \SI{6}{nm}$, making the TCI scheme appealing to such applications. Alternatively, bunch compression could result in sub-fs, $\gtrsim \SI{10}{kA}$ peak current bunches; similar schemes have been proposed for generating high peak power, attosecond-duration soft x-rays from plasma-based accelerators~\cite{emma2021terawatt}.

In summary, we have demonstrated a simple scheme to inject electrons directly into the wakefield driven by a pulse guided by an all-optical plasma channel. Dark-current-free, $\SI{1.2}{GeV}$ bunches with 4.5\% relative energy spread were generated with \SI{120}{TW} laser pulses guided in a \SI{110}{mm}-long HOFI plasma channel. Our measurements, supported by simulations, showed that high-quality electron bunches were only obtained when the drive pulse was well-aligned with the channel axis and focused close to the density down-ramp formed at the channel entrance, and that increasing the channel length to $L_\mathrm{d} \approx \SI{410}{mm}$ would yield \SI{3.65}{GeV} bunches, with relative slice energy spread below the per-mille level. In contrast, bunches injected via ionization were broadband and preferentially occurred when the drive was misaligned with the channel. This experiment is the first to exploit sculpting of the longitudinal and transverse density profile of all-optical plasma channels to control electron injection into a plasma channel accelerator stage. Further tailoring of the plasma channel could be achieved by manipulating the channel-forming pulse itself, or employing additional laser pulses. All techniques employed here are well-suited to $\gtrsim$ kHz repetition-rates, making this scheme promising for high-repetition-rate, compact radiation sources, including FELs.

The authors would like to acknowledge useful discussions with R\'emi Lehe. This work was supported by the UK Science and Technology Facilities Council (STFC UK) [grant numbers ST/R505006/1, ST/S505833/1 \& ST/V001655/1]; the Engineering and Physical Sciences Research Council [grant numbers EP/R513295/1 \& EP/V006797/1]; and the Central Laser Facility of the United Kingdom. This material is based upon work supported by the Air Force Office of Scientific Research under award number FA9550-18-1-7005. Computing resources provided by STFC Scientific Computing Department’s SCARF cluster. The CFD work used in this study was supported by funding from the CLF/EPAC (Extreme Photonics Applications Centre). The software used in this work was developed in part by the DOE NNSA- and DOE Office of Science-
supported Flash Center for Computational Science at the University of Chicago and the University of
Rochester.

This research was funded in whole, or in part, by EPSRC and STFC, which are Plan S funders. For the purpose of Open Access, the author has applied a CC BY public copyright licence to any Author Accepted Manuscript version arising from this submission.

\appendix*
\section{Appendix}
\subsection{Control over the density transition and channel characteristics} 
\label{sec:appendix}
It is important to be able to control the scale length of the longitudinal density transition, $\hat{L}_\mathrm{tr} = n_\mathrm{e0} / (d n_\mathrm{e} / d z )$, since this determines properties of the injected electron bunch. Furthermore, having independent control over the injector and accelerator section is vital for producing high quality electron bunches. Even though TCI uses a single laser pulse to both create the down-ramp and the plasma channel, it offers additional control and benefits over schemes where shocks are generated by interrupting the flow of gas \cite{tsai2018control, swanson2017control, fan2020gas}.

\begin{figure}[h!]
  \centering
  \includegraphics[width=46mm]{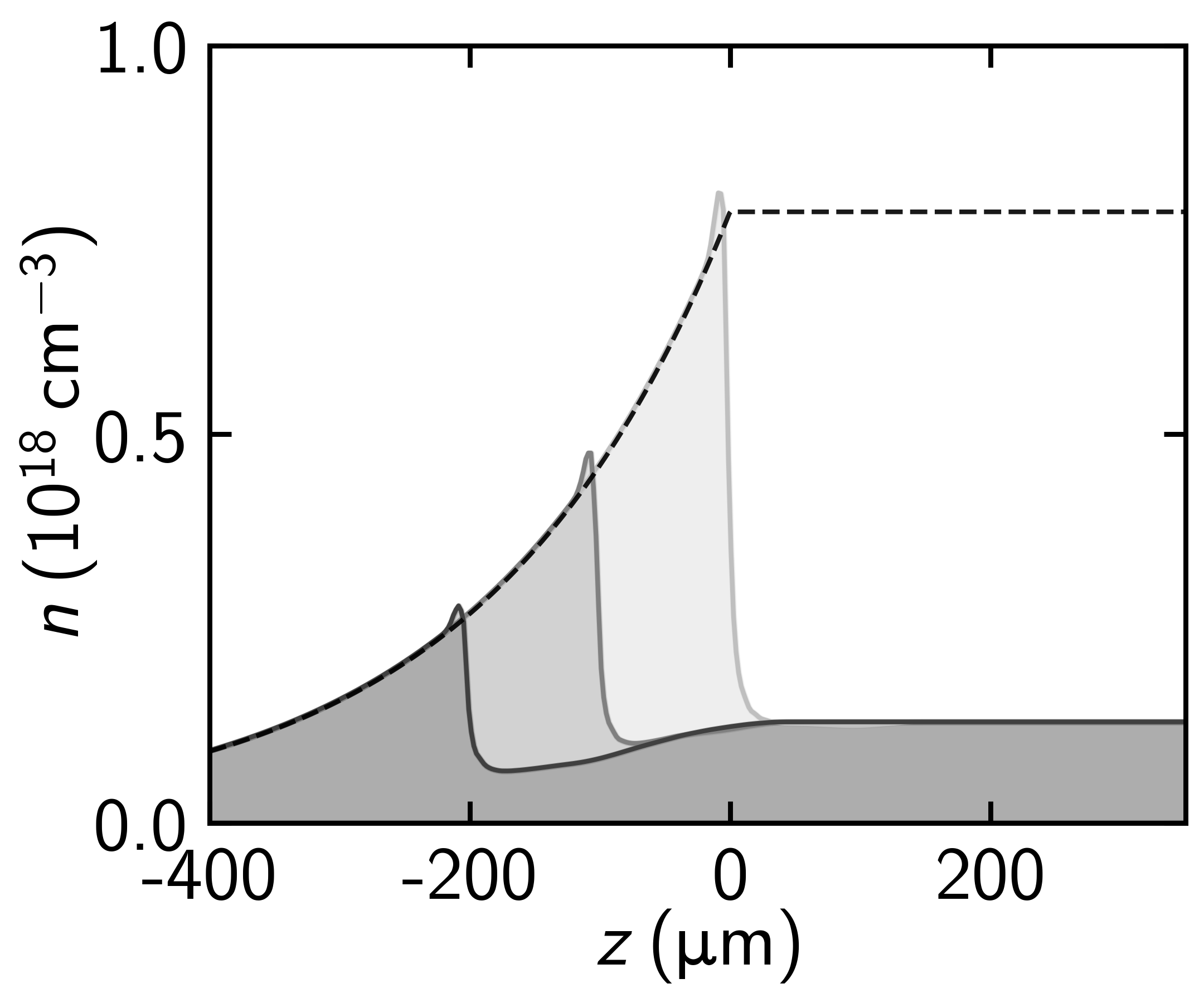}
  \caption{Results of fluid simulations of the longitudinal variation of the total density $n = n_\mathrm{e} + n_\mathrm{H}$ for three simulations with $T_\mathrm{e}(r, z - \delta z)$ with $\delta z = 0, -100, \SI{-200}{\micro m}$. The initial gas density profile, $n_{\mathrm{gas}}$, assumed in these simulations is represented by the black, dashed line.}
  \label{fig:app_ramp}
\end{figure}

With the TCI scheme, $\hat{L}_\mathrm{tr}$ is achieved by fixing the position of the gas cell entrance ($z = 0$) and varying the position of the start of the axicon focus $z_\mathrm{ch}$ within the gas plume. Figure~\ref{fig:app_ramp} illustrates this tunability, showing the longitudinal density profile $n(z) = (n_{\mathrm{e}} + n_{\mathrm{H}})$ calculated via MHD simulations~\cite{supp_mat} for a fixed initial gas density distribution and $T_\mathrm{e}(r,z - \delta z)$, with $\delta z = 0, -100, \SI{-200}{\micro m}$ where $T_\mathrm{e}(r,z)$ is shown in Fig.~\ref{fig:setup}(c). The black dashed line indicates $n_\mathrm{gas}$ at $\tau = 0$. It can be seen that the down-ramp length is $\approx \SI{70}{\micro m}$ and the ratio of the peak shock density to the on-axis density in the main part of the channel, $n_{\mathrm{shock}} / n_{\mathrm{e0}}$ can be varied from approximately 2 to 6, corresponding to a variation of $\hat{L}_\mathrm{tr} = \SIrange{70}{15}{\micro m}$, and thus allowing sufficient control over the injected electron bunch properties.

\begin{figure}[h!]
  \centering
  \includegraphics[width=60mm]{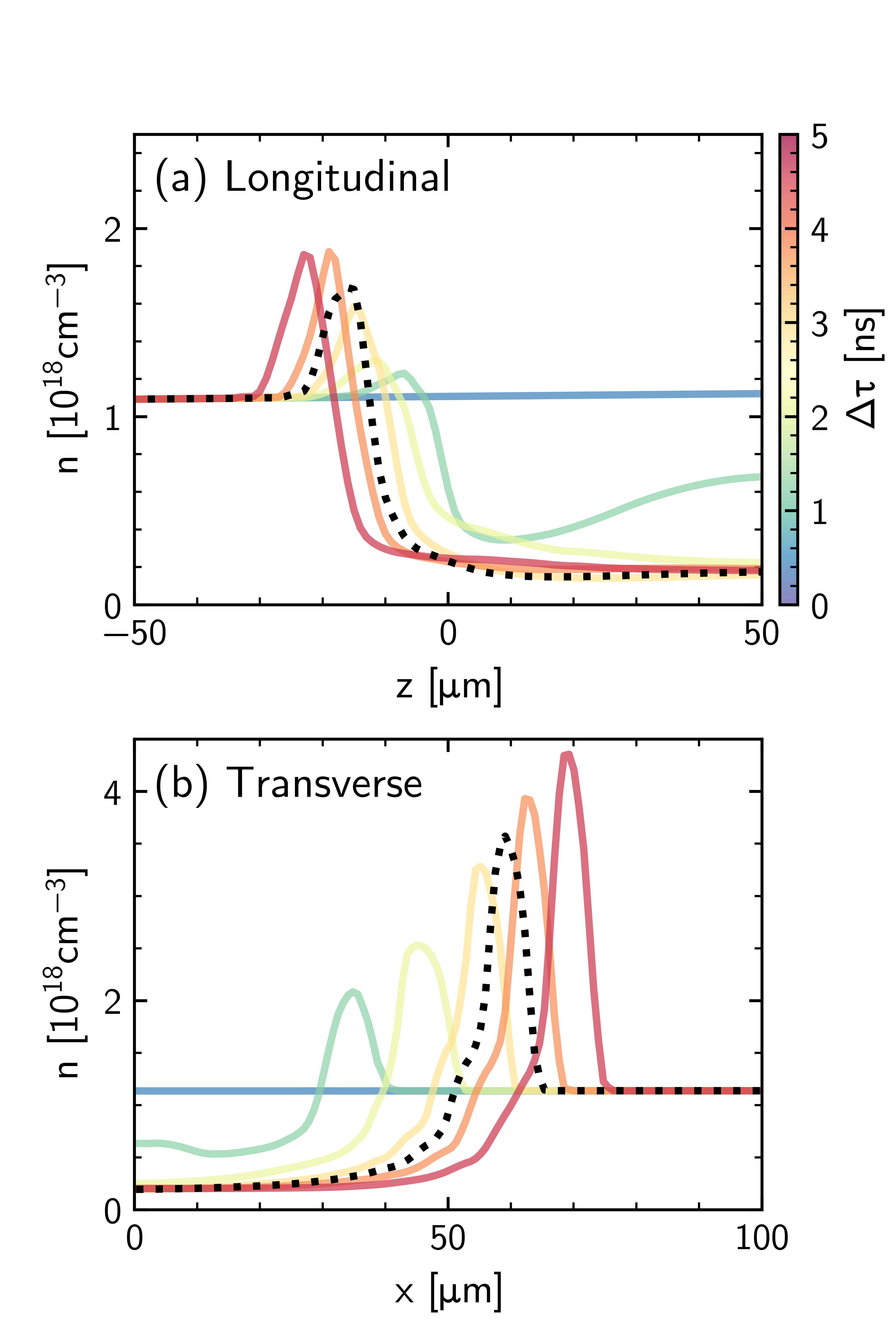}
  \caption{Results of fluid simulations showing the temporal evolution of the total density $n = n_\mathrm{e} + n_\mathrm{H}$ for: (a) the longitudinal expansion at the start of the channel; and (b) the transverse expansion of the bulk that forms the guiding structure. The delays plotted vary between $0$ to $\SI{5}{ns}$ in steps of \SI{1}{ns}. The delay used in the experiment, $\Delta \tau = \SI{3.5}{ns}$, is represented by the black, dotted line.}
  \label{fig:app_evo}
\end{figure}

Additionally, whilst the time delay $\Delta \tau$ between the two pulses determines the channel size and matching condition for the propagation section, it does not affect the properties of $\hat{L}_\mathrm{tr}$ for $\Delta \tau$ where guiding profiles exist. Figure~\ref{fig:app_evo} shows the evolution of both the longitudinal expansion of the leading edge of the channel and the transverse expansion of the bulk. The hemispherical expansion responsible for the injection section carries less energy-per-unit length compared to the cylindrical expansion of the channel body. After \SI{3}{ns} the shape of this transition is essentially unchanged, except for a small longitudinal shift. Over the same time period, the transverse expansion of the bulk continues, modifying the shape of the guiding channel and hence its matched spot size. This difference in expansion rate allows independent tuning of the injection region and guiding properties of the channel, without the need to modify the gas target. 

\newpage

\bibliography{references.bib}

\end{document}